\documentclass[aps,prc,preprint,amsmath,amssymb,showpacs,preprintnumbers,superscriptaddress]{revtex4-1}
\usepackage{CJK}
\usepackage{graphicx}
\usepackage{dcolumn}
\usepackage{bm}
\usepackage{color}
\usepackage{hyperref}

\allowdisplaybreaks[4]

\begin{document}

\title{Hamiltonian flow equations for a Dirac particle in large scalar and vector potentials}
\author{Z. X. Ren}
\affiliation{State Key Laboratory of Nuclear Physics and Technology, School of Physics, Peking University, Beijing 100871, China}

\author{P. W. Zhao}
\email{pwzhao@pku.edu.cn}
\affiliation{State Key Laboratory of Nuclear Physics and Technology, School of Physics, Peking University, Beijing 100871, China}

\begin{abstract}
An efficient solution of the Dirac Hamiltonian flow equations has been proposed through a novel expandsion with the inverse of the Dirac effective mass.
The efficiency and accuracy of this new expansion have been demonstrated by reducing a radial Dirac Hamiltonian with large scalar and vector potentials  to two nonrelativistic Hamiltonians corresponding to particles and antiparticles, respectively.
By solving the two nonrelativistic Hamiltonians, it is found that the exact solutions of the Dirac equation, for both particles and antiparticles, can be reproduced with a high accuracy up to only a few lowest order terms in the expansion.
This could help compare and bridge the relativistic and nonrelativistic nuclear energy density functional theories in the future.
\end{abstract}


\maketitle

\section{Introduction}

New experimental facilities with radioactive nuclear beams have helped us investigate the nuclear chart to the very limits of nuclear binding.
Considerable efforts on the theoretical side have been made to understand the dynamics of the nuclear many-body problem using microscopic methods.
Density functional theories play a very important role in this context~\cite{Bender2003Self}.
Because of the consideration of the Lorentz symmetry, the covariant density functional theory (CDFT) has attracted a lot of attentions in nuclear physics~\cite{meng2016relativistic}. It allows us to describe the spin-orbit coupling and the time-odd fields~\cite{Afanasjev1999PhysicsReport, Meng2013FT_TAC} in a consistent way; see e.g., Ref. \cite{meng2016relativistic} for details.

An essential ingredient of CDFT is to solve the relativistic Kohn-Sham equation with its effective single-particle potential, i.e., a Dirac equation with an attractive scalar potential and a repulsive vector potential.
The scalar and vector potentials are large, and they are around several hundred MeV.
The cancellation between the large attractive and repulsive potentials leads to a relatively weak potential felt by nucleons in the Fermi sea.
This reveals clearly the relativistic dynamics, rather than the relativistic kinematics, in describing the phenomena of low-energy nuclear structure.
For this reason, the nonrelativistic density functionals are also very successful if their parameters are carefully chosen.

Compared to the nonrelativistic description, the solution of the Dirac equation is more complicated than that of the Schr\"odinger equation because the Dirac spectrum is not bound from below. The direct application of conventional iterative methods to Dirac equation will meet several serious problems, such as variational collapse problem~\cite{ZhangIJMPE2010}.
To avoid these problems, one can adopt the inverse Hamiltonian method~\cite{hagino2010iterative}, which has been extended for three-dimensional (3D) lattice CDFT recently~\cite{tanimura20153d, REN2017Dirac3D, Ren2019C12LCS}, or solve the Schr\"{o}dinger-like equations for the upper and lower components of the Dirac spinors~\cite{BLUM1992364, ZhangIJMPE2010}.

In Ref.~\cite{bylev1998SRG}, a novel procedure for continuous unitary transformations, known as the similarity renormalization group (SRG) method,  was introduced to reduce the Dirac Hamiltonian to a quasidiagonal form, i.e., two noninteracting parts corresponding to the positive and negative energies, respectively.
The flow equations of the reduced nonrelativistic Hamiltonian are solved by an expansion in a series of $1/M$ ($M$ is the bare mass of the Dirac particle).
In contrast to the Schr\"{o}dinger-like equations, the Hermitian of the reduced nonrelativistic Hamiltonian here is guaranteed at every order of the expansion.
In recent years, this method has been adopted to investigate the pseudospin symmetries in nuclei with spherical~\cite{Guo2012SRG_spherical, Li2013SGR_PSS_Sph} and axial symmetries~\cite{Guo2014SRG_def, Li2015SRG_PSS_Def}, as well as the nuclear proton radioactivity~\cite{Zhao2014ProtonRadioactivity}.

It is interesting to develop a self-consistent CDFT by solving the Dirac equation with the SRG method, because this would allow a comparison between the nonrelativistic limit of the covariant density functionals and the nonrelativistic density functionals.
Such a comparison is motivated by the fact that the role and importance of the various terms in either covariant or nonrelativistic energy density functionals have not been completely understood so far.
Furthermore, different density-functional predictions exhibit systematic differences, which cannot yet be mapped onto the corresponding features of energy density functionals.
Solving the CDFT with the SRG method could help clarify these questions.

However, the proposed expansion of the Hamiltonian flow equations by $1/M$ in Ref.~\cite{bylev1998SRG} is not efficient enough to developing a self-consistent CDFT because of the existence of the large scalar potential in the Dirac equation.
The accuracy of the expansions up to the third order is still too large for a self-consistent solution, as shown in Refs.~\cite{Guo2012SRG_spherical,Guo2014SRG_def}.
In the present work, an efficient solution of the Dirac Hamiltonian flow equations
is proposed through a novel expansion with the inverse of the Dirac effective mass.
The efficiency and the accuracy of this proposed expansion are demonstrated for a radial Dirac Hamiltonian for spherical nuclei.

\section{Theoretical framework}
\subsection{General formalism}
In the framework of the CDFT, one needs to solve the Dirac equation with the Hamiltonian
\begin{equation}\label{Eq_dirac}
   H=\bm{\alpha}\cdot\bm{p}+\beta(M+S)+V,
\end{equation}
where $\bm{\alpha}$ and $\beta$ are the Dirac matrices, $M$ is the mass of nucleon, and $S$ and $V$ are the scalar and vector potentials, respectively.
With the SRG method~\cite{wegner1994SRG, bylev1998SRG}, the Hamiltonian $H$ can be transformed by a unitary operator $U(l)$ as
\begin{equation}\label{eq_UHU}
  H(l)=U(l)HU^\dag(l),~~H(0)=H,
\end{equation}
where $l$ is a flow parameter.
The flow equation can be obtained by calculating the derivative of $H(l)$ with respect to $l$,
\begin{equation}\label{eq_flow_eq}
  \frac{d H(l)}{d l}=[\eta(l), H(l)]
\end{equation}
with the generator,
\begin{equation}
  \eta(l)=\frac{d U(l)}{d l}U^\dag(l)=-\eta^\dag(l).
\end{equation}
In order to transform the Dirac Hamiltonian into a block-diagonal form, it is appropriate to choose the generator $\eta(l)$ in the form of
\begin{equation}\label{eq_eta}
   \eta(l)=[\beta, H(l)].
\end{equation}
Here, the generator $\eta(l)$ has the dimension of energy, and the flow parameter $l$ has the dimension of the inverse of energy.

To solve the flow equation Eq. \eqref{eq_flow_eq}, as in Ref. \cite{bylev1998SRG}, the Hamiltonian $H(l)$ is written as a sum of an even operator $\mathcal{E}(l)$ and an odd one $\mathcal{O}(l)$,
\begin{equation}\label{eq_H_O_D}
    H(l) = \mathcal{E}(l) + \mathcal{O}(l),
\end{equation}
where the even and odd operators are defined by the commutation relation with the $\beta$ matrix, i.e., $\mathcal{E}(l)\beta=\beta\mathcal{E}(l)$ and $\mathcal{O}(l)\beta=-\beta\mathcal{O}(l)$.
Through Eqs. \eqref{eq_eta} and \eqref{eq_H_O_D}, the flow equation \eqref{eq_flow_eq} can be split up into,
\begin{subequations}
   \begin{align}
     &\frac{d\mathcal{E}(l)}{d l}=4\beta\mathcal{O}^2(l),\label{eq_sub_El}\\
     &\frac{d\mathcal{O}(l)}{d l}=2\beta[\mathcal{O}(l), \mathcal{E}(l)],\label{eq_sub_Ol}
    \end{align}
\end{subequations}
with the initial conditions,
\begin{equation}
   \mathcal{E}(0)=\beta(M+S)+V,~~\mathcal{O}(0)=\bm{\alpha}\cdot\bm{p}.
\end{equation}

In contrast to Refs.~\cite{bylev1998SRG, Guo2012SRG_spherical}, where the flow equations \eqref{eq_sub_El} and \eqref{eq_sub_Ol} were solved by an expansion with the constant $1/M$,
here we solve the same equations by introducing a perturbative expansion in $1/\tilde{M}$ with $\tilde{M}=M+S$ being the Dirac effective mass, i.e.,
\begin{equation}\label{eq_expansion}
  \frac{1}{\tilde{M}}\mathcal{E}(l)=\sum_{k=0}^\infty\frac{1}{{\tilde{M}}^k}\mathcal{E}_k(l),\quad \quad
  \frac{1}{\tilde{M}}\mathcal{O}(l)=\sum_{k=1}^\infty\frac{1}{{\tilde{M}}^{k}}\mathcal{O}_k(l).
\end{equation}
It should be noted that $\tilde{M}$ does not always commute with $\mathcal{E}_k(l)$ or $\mathcal{O}_k(l)$ since it is not a constant but a function depending on the coordinate here.
Substituting Eq. \eqref{eq_expansion} into Eqs. \eqref{eq_sub_El} and \eqref{eq_sub_Ol}, one obtains the equations for the $n$-th order,
\begin{subequations}
  \begin{align}
   \frac{1}{\tilde{M}}\frac{d\mathcal{E}_n(l)}{d l}
    =&4\beta\sum_{k=1}^{n-1}{\tilde{M}}^{n-k-1}\mathcal{O}_k(l)\frac{1}{{\tilde{M}}^{n-k-1}}\mathcal{O}_{n-k}(l), \label{eq_sub_enl}\\
   \frac{1}{\tilde{M}}\frac{d\mathcal{O}_n(l)}{d l}
    =&-2\mathcal{O}_n(l)-2\frac{1}{\tilde{M}}\mathcal{O}_n(l)\tilde{M}+{\tilde{M}}^{n-2}2\beta\sum_{k=1}^{n-1}\left[\frac{1}{{\tilde{M}}^{k-1}}\mathcal{O}_k(l), \frac{1}{{\tilde{M}}^{n-k-1}}\mathcal{E}_{n-k}(l)\right].\label{eq_sub_onl}
  \end{align}
\end{subequations}
The solutions of Eqs. \eqref{eq_sub_enl} and \eqref{eq_sub_onl} read
\begin{subequations}
  \begin{align}
     \mathcal{E}_n(l)=&\mathcal{E}_n(0)+4\beta \tilde{M}\int_0^l d l'~ \sum_{k=1}^{n-1}{\tilde{M}}^{n-k-1}\mathcal{O}_k(l')\frac{1}{{\tilde{M}}^{n-k-1}}\mathcal{O}_{n-k}(l'),\\
     \mathcal{O}_n(l)=&e^{-2\tilde{M} l}\mathcal{O}_n(0)e^{-2\tilde{M}l}\nonumber\\
     &+e^{-2\tilde{M} l}\left\{2\beta{\tilde{M}}^{n-1}\int_0^l d l'~e^{2\tilde{M}l'}\sum_{k=1}^{n-1}\left[\frac{1}{{\tilde{M}}^{k-1}},\mathcal{O}_k(k'),\frac{1}{{\tilde{M}}^{n-k-1}}\mathcal{E}_{n-k}(l)\right]e^{2\tilde{M} l'}\right\}e^{-2\tilde{M} l},
  \end{align}
\end{subequations}
with the initial conditions,
\begin{subequations}
  \begin{align}
    &\mathcal{E}_0(0)=\beta,\quad \mathcal{E}_1(0)=V,\quad \mathcal{E}_n(0)=0~\textmd{if}~n\geq2,\\
    &\mathcal{O}_1(0)=\bm{\alpha}\cdot\bm{p},\quad \mathcal{O}_n(0)=0~\textmd{if}~n\geq2.\label{eq_inital_O1}
  \end{align}
\end{subequations}
One can verify that $\mathcal{O}_n(l)$ exponentially goes to zero while $\mathcal{E}_n(l)$ is finite as $l\rightarrow\infty$,
\begin{equation}\label{eq_sum_e_inf}
  \mathcal{E}(\infty)=\tilde{M}\mathcal{E}_0(\infty)+\mathcal{E}_1(\infty)+\frac{1}{\tilde{M}}\mathcal{E}_2(\infty)+\frac{1}{{\tilde{M}}^2}\mathcal{E}_3(\infty)+\frac{1}{{\tilde{M}}^3}\mathcal{E}_4(\infty)+\cdots,
\end{equation}
with the terms up to ${1}/{\tilde M^3}$ order read,
\begin{subequations}
  \begin{align}
    &\tilde{M}\mathcal{E}_0(\infty)+\mathcal{E}_1(\infty)=\beta \tilde{M} + V,\label{eq_e0_transform}\\
    &\frac{1}{\tilde{M}}\mathcal{E}_2(\infty)=\beta\left\{\mathcal{O}_1(0)\frac{1}{2\tilde{M}}\mathcal{O}_1(0) - \frac{1}{8{\tilde{M}}^2}[\mathcal{O}_1(0), [\mathcal{O}_1(0), \tilde{M}]]+\frac{3}{16{\tilde{M}}^3}[\mathcal{O}_1(0), \tilde{M}]^2 \right\},\\
    &\frac{1}{{\tilde{M}}^2}\mathcal{E}_3(\infty)=\frac{1}{8{\tilde{M}}^3}[\mathcal{O}_1(0),\tilde{M}][\mathcal{O}_1(0),\mathcal{E}_1(0)]-\frac{1}{8{\tilde{M}}^2}[\mathcal{O}_1(0),[\mathcal{O}_1(0),\mathcal{E}_1(0)]],\\
      &\frac{1}{{\tilde{M}}^3}\mathcal{E}_4(\infty)=-\beta\mathcal{O}_1^2(0)\frac{1}{8{\tilde{M}}^3}\mathcal{O}_1^2(0)-\frac{1}{16{\tilde{M}}^3}\beta[\mathcal{O}_1(0),\mathcal{E}_1(0)]^2\nonumber\\
      &\qquad\qquad\qquad~~~-\beta\mathcal{O}_1(0)\left\{\frac{13}{32}\frac{[\mathcal{O}_1(0),\tilde{M}]^2}{{\tilde{M}}^5}-\frac{15}{64}\frac{[\mathcal{O}_1(0),[\mathcal{O}_1(0),\tilde{M}]]}{{\tilde{M}}^4}\right\}\mathcal{O}_1(0)\nonumber\\
      &\qquad\qquad\qquad~~~-\beta\left\{\frac{17}{128}\frac{[\mathcal{O}_1(0),\tilde{M}]^2}{{\tilde{M}}^5}-\frac{9}{128}\frac{[\mathcal{O}_1(0),[\mathcal{O}_1(0),\tilde{M}]]}{{\tilde{M}}^4}\right\}\mathcal{O}_1^2(0)\nonumber\\
      &\qquad\qquad\qquad~~~-\beta\mathcal{O}_1^2(0)\left\{\frac{17}{128}\frac{[\mathcal{O}_1(0),\tilde{M}]^2}{{\tilde{M}}^5}-\frac{9}{128}\frac{[\mathcal{O}_1(0),[\mathcal{O}_1(0),\tilde{M}]]}{{\tilde{M}}^4}\right\}\nonumber\\
      &\qquad\qquad\qquad~~~-\beta\left\{\frac{87}{512}[\mathcal{O}_1(0),[\mathcal{O}_1(0),\tilde{M}]]\frac{[\mathcal{O}_1(0),\tilde{M}]}{{\tilde{M}}^5}-\frac{735}{2048}\frac{[\mathcal{O}_1(0),\tilde{M}]^3}{{\tilde{M}}^6}\right\}\mathcal{O}_1(0)\nonumber\\
      &\qquad\qquad\qquad~~~+\beta\mathcal{O}_1(0)\left\{\frac{87}{512}\frac{[\mathcal{O}_1(0),\tilde{M}]}{{\tilde{M}}^5}[\mathcal{O}_1(0),[\mathcal{O}_1(0),\tilde{M}]]-\frac{735}{2048}\frac{[\mathcal{O}_1(0),\tilde{M}]^3}{{\tilde{M}}^6}\right\}\nonumber\\
      &\qquad\qquad\qquad~~~-\frac{27}{512}\beta\frac{[\mathcal{O}_1(0),[\mathcal{O}_1(0),\tilde{M}]]^2}{{\tilde{M}}^5}+\frac{489}{2048}\beta\frac{[\mathcal{O}_1(0),\tilde{M}]^2[\mathcal{O}_1(0),[\mathcal{O}_1(0),\tilde{M}]]}{{\tilde{M}}^6}\nonumber\\
      &\qquad\qquad\qquad~~~-\frac{293}{1024}\beta\frac{[\mathcal{O}_1(0),\tilde{M}]^4}{{\tilde{M}}^7}. \label{eq_e4_transform}
  \end{align}
\end{subequations}
Here, we define the powering counting of the commutators $[\,\, , \tilde{M} ]$ as the same order of $\tilde{M}$.
One can easily see that the Hermitian is automatically satisfied at each order of the expansion.

Finally, the $\mathcal{E}(\infty)$ can thus be written as a block-diagonal form,
\begin{equation}\label{eq_block_E_inf}
   \mathcal{E}(\infty)=
   \begin{pmatrix}
     H^F+M&0\\
     0&H^D-M,
   \end{pmatrix}
\end{equation}
where $H^F$ and $H^D$ describe the Dirac particles and antiparticles, respectively.
By assuming vanished scalar $S$ and vector $V$ potentials, one can find from Eqs. \eqref{eq_e0_transform}-\eqref{eq_e4_transform} that the Hamiltonian $H^F$ in Eq. \eqref{eq_block_E_inf} reads
\begin{equation}
  H^F=p^2/2M-p^4/8M^3+\cdots.
\end{equation}
This is nothing but the nonrelativistic expansion of the relativistic kinetic energy $\sqrt{p^2+M^2}-M$, so the Hamiltonian $H^F$ has the basic property of a nonrelativistic Hamiltonian, i.e., its spectrum is bound from below.

\subsection{Spherical case}
For a system with spherical symmetry, one can write the radial Dirac equation as
\begin{equation}\label{eq_Dirac_sph}
   \begin{pmatrix}
     V+S+M   &-\frac{d}{d r}+\frac{\kappa}{r}\\
     \frac{d}{d r}+\frac{\kappa}{r}&V-S-M
   \end{pmatrix} \begin{pmatrix} G(r)\\F(r)\end{pmatrix}=E\begin{pmatrix} G(r)\\F(r)\end{pmatrix},
\end{equation}
where $\kappa=(-1)^{j+l+1/2}(j+1/2)$, $E$ is the single-particle energy, and $G(r)$ and $F(r)$ are the upper and
lower components of the Dirac spinors, respectively. The corresponding initial condition for $\mathcal{O}_1(0)$ [see Eq. \eqref{eq_inital_O1}] is thus reduced as
\begin{equation}
   \mathcal{O}_1(0)=
   \begin{pmatrix}
      0  &-\frac{d}{d r}+\frac{\kappa}{r}\\
      \frac{d}{d r}+\frac{\kappa}{r} &0
   \end{pmatrix}.
\end{equation}
Accordingly, the obtained $H^F$ and $H^D$ [see Eq. \eqref{eq_block_E_inf}], up to the $1/{\tilde{M}}^3$ order, respectively, read
\begin{subequations}\label{eq_nr_HF}
  \begin{align}
    H_0^F=&\,V+S,\label{eq_nr_HF_0}\\
    H_1^F=&-\frac{d}{d r}\frac{1}{2\tilde{M}}\frac{d}{d r}+\frac{1}{2\tilde{M}}\frac{\kappa(\kappa+1)}{r^2}+\frac{S'}{4{\tilde{M}}^2}\frac{\kappa}{r}+\frac{S''}{8{\tilde{M}}^2}-\frac{3S'^2}{16{\tilde{M}}^3},\label{eq_nr_HF_1}\\
    H_2^F=&-\frac{V'}{4{\tilde{M}}^2}\frac{\kappa}{r}+\frac{1}{8{\tilde{M}}^2}V''-\frac{1}{8{\tilde{M}}^3}S'V',\label{eq_nr_HF_2}\\
    H_3^F=&-p^2\frac{1}{8{\tilde{M}}^3}p^2+\frac{V'^2}{16{\tilde{M}}^3}\nonumber\\
     &-\left(-\frac{d}{d r}+\frac{\kappa}{r}\right)\left(-\frac{13}{32}\frac{S'^2}{{\tilde{M}}^5}+\frac{15}{64}\frac{S''}{{\tilde{M}}^4}+\frac{15}{32}\frac{S'}{{\tilde{M}}^4}\frac{\kappa}{r}\right)\left(\frac{d}{d r}+\frac{\kappa}{r}\right)\nonumber\\
     &-\left(-\frac{17}{128}\frac{S'^2}{{\tilde{M}}^5}+\frac{9}{128}\frac{S''}{{\tilde{M}}^4}-\frac{9}{64}\frac{S'}{{\tilde{M}}^4}\frac{\kappa}{r}\right)p^2-p^2\left(-\frac{17}{128}\frac{S'^2}{{\tilde{M}}^5}+\frac{9}{128}\frac{S''}{{\tilde{M}}^4}-\frac{9}{64}\frac{S'}{{\tilde{M}}^4}\frac{\kappa}{r}\right)\nonumber\\
     &-\left(\frac{87}{512}\frac{S'S''}{{\tilde{M}}^5}-\frac{87}{256}\frac{S'^2}{{\tilde{M}}^5}\frac{\kappa}{r}-\frac{735}{2048}\frac{S'^3}{{\tilde{M}}^6}\right)\left(\frac{d}{d r}+\frac{\kappa}{r}\right)\nonumber\\
     &-\left(-\frac{d}{d r}+\frac{\kappa}{r}\right)\left(\frac{87}{512}\frac{S'S''}{{\tilde{M}}^5}-\frac{87}{256}\frac{S'^2}{{\tilde{M}}^5}\frac{\kappa}{r}-\frac{735}{2048}\frac{S'^3}{{\tilde{M}}^6}\right)\nonumber\\
     &-\frac{27}{512}\frac{1}{{\tilde{M}}^5}\left(S''-2S'\frac{\kappa}{r}\right)^2+\frac{489}{2048}\frac{1}{{\tilde{M}}^6}\left(S''S'^2-2S'^3\frac{\kappa}{r}\right)-\frac{293}{1024}\frac{S'^4}{{\tilde{M}}^7},
  \end{align}
\end{subequations}
and
\begin{subequations}\label{eq_nr_HD}
  \begin{align}
    H_0^D=&V-S,\\
    H_1^D=&\frac{d}{d r}\frac{1}{2\tilde{M}}\frac{d}{d r}-\frac{1}{2\tilde{M}}\frac{\kappa(\kappa-1)}{r^2}+\frac{S'}{4{\tilde{M}}^2}\frac{\kappa}{r}-\frac{S''}{8{\tilde{M}}^2}+\frac{3S'^2}{16{\tilde{M}}^3},\\
    H_2^D=&\frac{V'}{4{\tilde{M}}^2}\frac{\kappa}{r}+\frac{1}{8{\tilde{M}}^2}V''-\frac{1}{8{\tilde{M}}^3}S'V',\\
    \mathcal{H}_3^D=&p^2\frac{1}{8{\tilde{M}}^3}p^2-\frac{V'^2}{16{\tilde{M}}^3}\nonumber\\
     &-\left(\frac{d}{d r}+\frac{\kappa}{r}\right)\left(\frac{13}{32}\frac{S'^2}{{\tilde{M}}^5}-\frac{15}{64}\frac{S''}{{\tilde{M}}^4}+\frac{15}{32}\frac{S'}{{\tilde{M}}^4}\frac{\kappa}{r}\right)\left(-\frac{d}{d r}+\frac{\kappa}{r}\right)\nonumber\\
     &-\left(\frac{17}{128}\frac{S'^2}{{\tilde{M}}^5}-\frac{9}{128}\frac{S''}{{\tilde{M}}^4}-\frac{9}{64}\frac{S'}{{\tilde{M}}^4}\frac{\kappa}{r}\right)p^2-p^2\left(\frac{17}{128}\frac{S'^2}{{\tilde{M}}^5}-\frac{9}{128}\frac{S''}{{\tilde{M}}^4}-\frac{9}{64}\frac{S'}{{\tilde{M}}^4}\frac{\kappa}{r}\right)\nonumber\\
     &-\left(\frac{87}{512}\frac{S'S''}{{\tilde{M}}^5}+\frac{87}{256}\frac{S'^2}{{\tilde{M}}^5}\frac{\kappa}{r}-\frac{735}{2048}\frac{S'^3}{{\tilde{M}}^6}\right)\left(-\frac{d}{d r}+\frac{\kappa}{r}\right)\nonumber\\
     &-\left(\frac{d}{d r}+\frac{\kappa}{r}\right)\left(\frac{87}{512}\frac{S'S''}{{\tilde{M}}^5}+\frac{87}{256}\frac{S'^2}{{\tilde{M}}^5}\frac{\kappa}{r}-\frac{735}{2048}\frac{S'^3}{{\tilde{M}}^6}\right)\nonumber\\
     &+\frac{27}{512}\frac{1}{{\tilde{M}}^5}\left(S''+2S'\frac{\kappa}{r}\right)^2-\frac{489}{2048}\frac{1}{{\tilde{M}}^6}\left(S''S'^2+2S'^3\frac{\kappa}{r}\right)+\frac{293}{1024}\frac{S'^4}{{\tilde{M}}^7}.
  \end{align}
\end{subequations}
Here, we define $p^2=-\frac{d^2}{d r^2}+\frac{\kappa(\kappa+1)}{r^2}$ for $H^F$, while $p^2=-\frac{d^2}{d r^2}+\frac{\kappa(\kappa-1)}{r^2}$ for $H^D$.
The primes and the double primes denote the first- and second-order derivatives with respect to $r$, respectively.

\section{Numerical details}
In the following, the newly proposed solutions for the flow equations are applied to solve the Dirac equation with spherical Woods-Saxon potentials
\begin{equation}
  V+S =\frac{\Sigma_0}{1+e^{\frac{r-r_0}{a_0}}}, \quad  V-S=\frac{\Delta_0}{1+e^{\frac{r-r_0}{a_0}}},
\end{equation}
where the parameters of the potentials are the same as those in Ref. \cite{Guo2012SRG_spherical}, i.e., $\Sigma_0=-66.0$~MeV, $\Delta_0=650.0$~MeV, $r_0=7.0$~fm, and $a_0=0.6$~fm, which are determined by fitting the single-neutron energies of $^{208}$Pb~\cite{Guo2005PSS_Pb208}.
The nucleon mass in the Dirac equation \eqref{eq_Dirac_sph} is taken as $M=939$~MeV.

The reduced Hamiltonians $H^F$ and $H^D$ [see Eqs. \eqref{eq_nr_HF} and \eqref{eq_nr_HD}] are solved in a large set of spherical harmonic oscillator bases. The obtained results are compared with the corresponding exact solutions of the Dirac equation, which are obtained by the shooting method~\cite{meng1998NPA} with the box size $R=30$ fm and the mesh size $0.05$ fm.

\section{Results and discussion}
\begin{figure}[!htbp]
  \centering
  \includegraphics[width=0.45\textwidth]{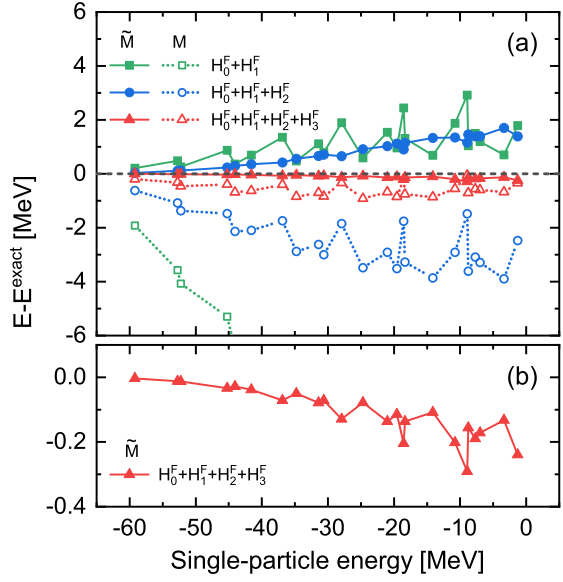}\\
  \caption{(Color online) The deviations of single-particle energies given by the SRG method from the exact ones as a function of single-particle energy.
  The solid and dotted lines represent the results obtained with the $\tilde{M}$ and $M$ expansions, respectively.
  The squares, circles and triangles represent respectively the energy deviations up to $1/\tilde{M}$, $1/{\tilde{M}}^2$, and $1/{\tilde{M}}^3$ orders for the $\tilde{M}$ expansion and $1/M$, $1/{M}^2$, and $1/{M}^3$ orders for the $M$ expansion.
  Panel (b) shows the difference to the exact results for the $\tilde{M}$ expansion up to $1/\tilde{M}^3$ order on a smaller scale.
  }\label{fig1}
\end{figure}

In Fig. \ref{fig1}, the differences between single-particle energies given by the SRG method and the exact ones are presented.
For comparison, the SRG results obtained by solving the flow equations with both the $\tilde{M}$ and $M$ expansions are shown.
One can immediately see that the results of the $\tilde{M}$ expansion exhibit a much faster convergence than those of the $M$ expansion.
At each order, the $\tilde{M}$ expansion provides more accurate solutions.
In particular, up to the $1/{\tilde{M}}^3$ order, the deviations of single-particle energies from the exact ones are less than $0.1$ MeV for the well bound levels, and they become slightly larger to about 0.2 MeV for the weakly bound levels [see Fig. \ref{fig1} (b)].
This is understandable because the weakly bound levels are usually sensitive to the higher momentum components and, thus, require the expansion terms at higher orders.
It is also interesting to note that the $\tilde{M}$ expansion up to $1/\tilde{M}^3$ order can reproduce the exact total energy,
here in terms of the sum of the single-particle energies of the lowest 126 levels, with a relative deviation of about $0.5\%$.

\begin{figure}[!htbp]
  \centering
  \includegraphics[width=0.45\textwidth]{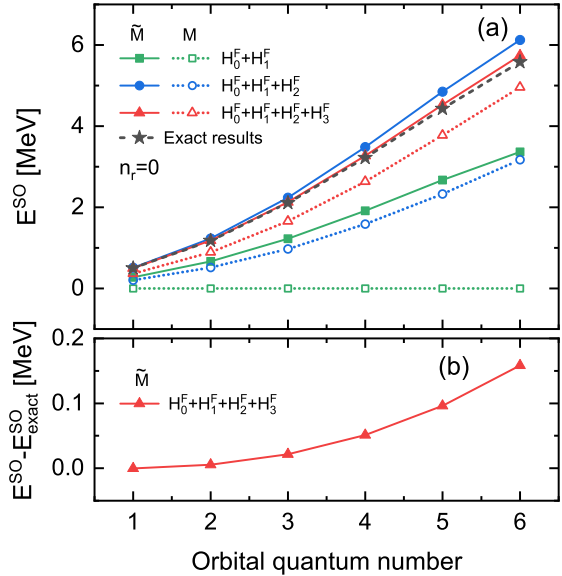}\\
  \caption{(Color online) (a): The spin-orbit splitting energies $E^{\rm SO}=E_{lj=l-1/2}-E_{lj=l+1/2}$ given by the SRG method and the exact solutions as a function of the orbital quantum number for each pair of the spin-orbit partners with the radial node number $n_r=0$.
  (b): The difference of $E^{\rm SO}$ to the exact results for the $\tilde{M}$ expansion up to $1/\tilde{M}^3$ order as a function of the orbital quantum number.
  }\label{fig2}
\end{figure}

One of the main features of the relativistic framework is the natural inclusion of the spin-orbit interactions.
In Fig. \ref{fig2} (a), the spin-orbit splitting energies $E^{\rm SO}=E_{lj=l-1/2}-E_{lj=l+1/2}$ given by the SRG method and the exact solutions are shown as a function of the orbital quantum number for each pair of the spin-orbit partners with the radial node number $n_r=0$.
It is seen that the results up to the second order $H_2^F$ in the $\tilde{M}$ expansions (solid circles) have already achieved a much better accuracy than the corresponding third-order results in the $M$ expansions (open triangles).
In fact, one can readily identify a spin-orbit potential from the expressions of $H_0^F$, $H_1^F$, and $H_2^F$ in Eqs.~\eqref{eq_nr_HF_0}-\eqref{eq_nr_HF_2}, i.e.,
\begin{equation}
   -\frac{V'-S'}{4{\tilde{M}}^2}\frac{\kappa}{r}.
\end{equation}
This reveals that this potential contributes the most part of the spin-orbit splitting energies.
By further including $H^F_3$ for the higher order contributions, the obtained spin-orbit splitting energies can excellently reproduce the exact results,
and the corresponding deviations from the exact results are smaller than 0.2 MeV [see Fig. \ref{fig2} (b)].
Therefore, one can in principle obtain the high-order corrections for the spin-orbit potential by analyzing the corresponding terms in $H_3^F$.

\begin{figure}[!ht]
  \centering
  \includegraphics[width=0.45\textwidth]{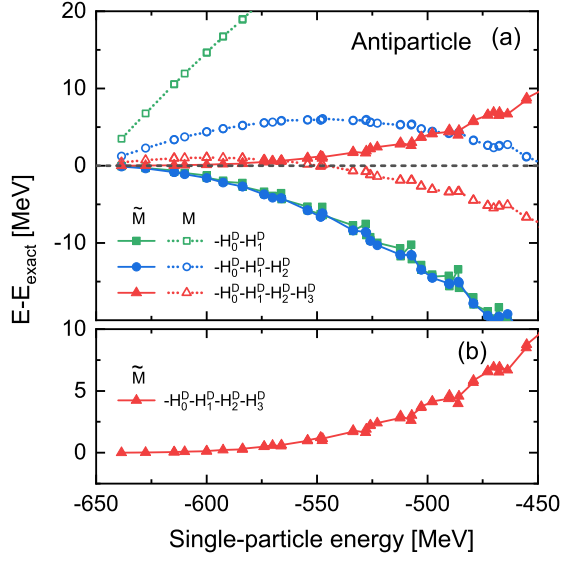}\\
  \caption{(Color online) Same as Fig. \ref{fig1} but for Dirac antiparticles.
  }\label{fig3}
\end{figure}

Apart from the solutions for Dirac particles, an exact solution of the Dirac equation gives rise to the results for antiparticles as well.
In Fig. \ref{fig3}, the differences between the antiparticle energies given by the SRG method and the exact ones are presented.
Similar to the spectrum of the Dirac particles, the antiparticle energies given by the SRG method with the $\tilde{M}$ expansion can reproduce the exact solutions in a very efficient way.
There are distinct features, however.
The second-order corrections $H^D_2$ are subtle in the $\tilde{M}$ expansions, while the third-order ones $H^D_3$ are remarkable.
This can be attributed to fact that the $H^D_2$ corrections mainly introduce the spin-orbit splittings, which are found to be small for the antiparticle spectrum in nuclei, known as the spin symmetry~\cite{zhou2003spin}.
For the remarkable $H^D_3$ corrections, the main contribution is from the term $\displaystyle p^2\frac{1}{8\tilde M^3}p^2$, which brings considerable high-order corrections to the kinetic energy.

Up to the $H^D_3$ corrections, the $\tilde{M}$ expansion provides a better agreement with the exact solutions than the $M$ expansion method for deeply bound levels with the deviations less than 0.1 MeV,
while the deviations are gently growing up for weakly bound levels and are close to 10 MeV with single-particle energy approaching $-450$ MeV [see Fig. \ref{fig3} (b)], similar to the case of the particle energies (see Fig. \ref{fig1}).
However, the antiparticle energies given by the $M$ expansion exhibit a visible overestimation for deeply bound levels but a considerable underestimation for weakly bound levels. Therefore, for several certain levels in between, an accident match between the $M$ expansion results and the exact solutions appears.

\begin{figure}[!ht]
  \centering
  \includegraphics[width=0.45\textwidth]{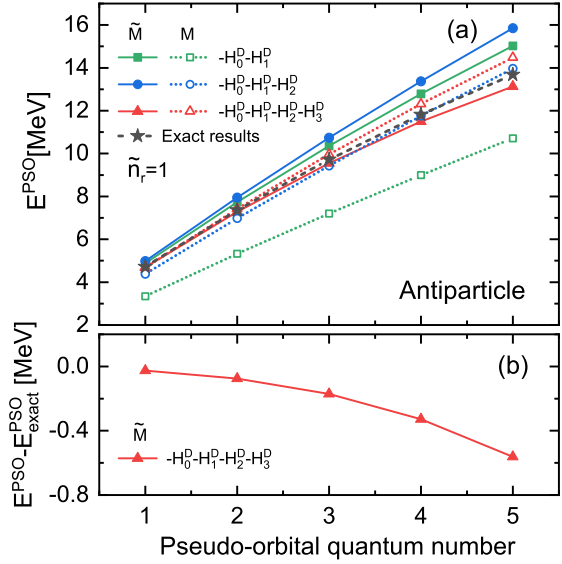}
  \caption{(Color online) Same as Fig. \ref{fig3}, but for the pseudospin-orbit splittings $E^{\rm PSO}=E_{\tilde{l}j=\tilde{l}-1/2}-E_{\tilde{l}j=\tilde{l}+1/2}$ with the pseudoradial node number $\tilde{n}_r=1$ as a function of the pseudo-orbital quantum number.
  }\label{fig4}
\end{figure}

Though the spin-orbit splittings are very small in the antiparticle spectrum, the corresponding pseudospin-orbit splittings are significant.
As in Ref.~\cite{zhou2003spin}, the pseudoquantum $\{\tilde{n}_r\tilde{l}j\}$ can be introduced for each antiparticle state $\{n_rlj\}$,
\begin{equation}
   \begin{cases}
     \tilde{n}_r=n_r,~~\tilde{l}=l+1,&\mbox{for}~j=l+1/2;\\
     \tilde{n}_r=n_r+1,~~\tilde{l}=l-1,&\mbox{for}~j=l-1/2.\\
   \end{cases}
\end{equation}
In Fig.~\ref{fig4} (a), the pseudospin-orbit splitting energies $E^{\rm PSO}=E_{\tilde{l}j=\tilde{l}-1/2}-E_{\tilde{l}j=\tilde{l}+1/2}$ are shown as a function of the  pseudo-orbital quantum number $\tilde{l}$ for the levels with the pseudo-radial node number $\tilde{n}_r=1$.
Up to $H^D_1$, the newly proposed $\tilde{M}$ expansions can reproduce the exact results satisfactorily, while the $M$ expansion results are far away from the exact ones.
By taking into account higher order corrections including $H_2^D$ and $H_3^D$, the exact solutions of the pseudospin-orbit splitting energies can be well reproduced
by both $\tilde{M}$ and $M$ expansions, in particular, for deeply bound levels.
For the $\tilde{M}$ expansion up to $1/\tilde{M}^3$ order, the deviations to the exact results are smaller than 0.6 MeV up to $\tilde{l}=5$ [see Fig. \ref{fig4} (b)].
This indicates that in contrast to the spin-orbit potentials, the pseudospin-orbit potentials are not grabbed very effectively in the lowest several orders of both expansions.
In fact, it is known that the nuclear pseudospin-orbit splitting is a consequence of complex dynamical cancellations of many different terms~\cite{Guo2012SRG_spherical,liang2015hidden}.

\begin{figure}[!ht]
  \centering
  \includegraphics[width=0.45\textwidth]{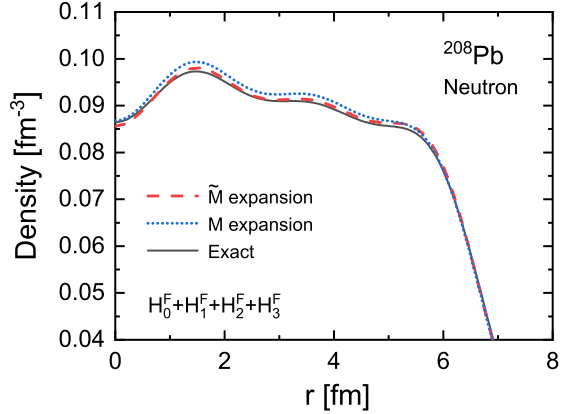}
  \caption{(Color online) The total density of the lowest 126 levels as a function of the radial coordinate $r$.
  The dashed and dotted lines represent results of the SRG method with the $\tilde{M}$ expansions up to $1/{\tilde{M}}^3$ order and the $M$ expansions up to $1/{M}^3$ order, respectively.
  The solid line denotes the exact results.
  }\label{fig5}
\end{figure}

Apart from the single-particle energies, the accuracy of the wave functions should also be examined.
This is illustrated in Fig. \ref{fig5}, where the total density of the lowest 126 single-particle levels are shown as a function of the radial coordinate $r$.
It should be noted that the accuracy of the total density plays a crucial role in the future implement of the SRG method to CDFT, because even a slight variation of the density could be enlarged by the self-consistent iteration procedure and lead to substantial discrepancies for the final solutions.
Here, since the considered Woods-Saxon potential is obtained for $^{208}$Pb, the total densities of the 126 neutrons are depicted.
In comparison with the $M$ expansions, one can see that results given by the $\tilde{M}$ expansions are closer to the exact ones.
The root mean square radii associated with these density profiles are respectively 5.61 fm, 5.59 fm, and 5.58 fm for the exact solution, the $\tilde{M}$ expansion, and $M$ one.
Both expansions reproduce the exact solution well, but the $\tilde{M}$ expansion is slightly more accurate.
These excellent agreements also pave a way for the future implementation of the $\tilde{M}$ expansions in the self-consistent CDFT.

Finally, it is worthwhile to mention that in parallel with our work, a so-called reconstituted SRG method has been proposed very recently for solving the radial Dirac equation by using the resummation technique~\cite{Guo2019SRG}.
It is found that all the terms in the reduced Hamiltonian from the reconstituted SRG can be obtained in the present work by directly solving the flow equations with the newly proposed $\tilde{M}$ expansions.
Because of its direct connection to the flow equations, the Hermitian of the terms at each order is satisfied automatically.
Moreover, the present derivations can be straightforwardly applied to reduce the Dirac equation with deformed potentials and time-odd fields~\cite{Zhao2018Spectroscopies}.

\section{Summary}
In summary, an efficient solution of the Dirac Hamiltonian flow equations has been proposed through a novel expandsion with the inverse of the Dirac effective mass.
With the new solutions of the Dirac Hamiltonian flow equations, one can reduce the Dirac Hamiltonian with large scalar and vector potentials into two nonrelativistic Hamiltonians corresponding to particles and antiparticles, respectively.
Taking the radial Dirac Hamiltonian with a spherical Woods-Saxon potential as an example, the efficiency and the accuracy of the expanding terms in the reduced nonrelativistic Hamiltonians have been demonstrated
by comparing them with the exact solutions of the Dirac Hamiltonian.
By solving the two nonrelativistic Hamiltonians, it is found that the exact solutions of the Dirac equation, for both particles and antiparticles, can be reproduced with a high accuracy up to only a few lowest order terms in the expansion.
In particular, for the large spin-orbit splittings of in the particle spectrum, the new expansion can achieve a much more efficient convergence in comparison with the previous expansion scheme.

A self-consistent calculation for the covariant density functional theory with the present new expansion scheme is rather straightforward,
but in addition to vector density, one also needs to calculate the scalar density $\rho_s$, which involves the transformation of the Dirac matrix $\beta$ under the unitary operator $U(l)$ in Eq. \eqref{eq_UHU}, $\beta(l)=U^\dag(l)\beta U(l)$.
The $\beta(l)$ can be solved via the equation of motion $\frac{d\beta(l)}{dl}=2[\beta\mathcal{O}(l),\beta(l)]$ with the initial condition $\beta(0)=\beta$.
This could help to compare and bridge the relativistic and nonrelativistic nuclear energy density functional theories in the future.
Works following this direction are in progress.

\begin{acknowledgments}
This work was partly supported by the National Key R\&D Program of China (Contracts No. 2018YFA0404400 and No. 2017YFE0116700) and
the National Natural Science Foundation of China (Grants No. 11621131001, No. 11875075, No. 11935003, and No. 11975031).
\end{acknowledgments}


%

\end{document}